\documentclass[journal]{IEEEtran}
\usepackage{amsmath}
\usepackage{graphicx}
\usepackage{fancyhdr}
\usepackage[mathscr]{euscript}
\usepackage{amsmath}
\usepackage{amsthm}
\usepackage[ruled,vlined]{algorithm2e}
\usepackage{epstopdf}
\usepackage{float}
\usepackage{amssymb}
\usepackage{subcaption}
\usepackage{url}

\begin{document}
%
\title{Optimal Demand Side Management by Distributed and Secured Energy Commitment Framework}
%
%
%

\author{Shantanu~Chakraborty,~and
 Toshiya~Okabe\thanks{Authors are with Smart Energy Research Laboratories, NEC Corporation, Japan.}%
}

\maketitle

\begin{abstract}
This paper introduces a demand-side distributed and secured energy commitment framework and operations for a Power Producer and Supplier (PPS) in deregulated environment. Due to the diversity of geographical location as well as customers' energy profile coupled with high number of customers, managing energy transactions and resulting energy exchanges are challenging for a PPS. The envisioned PPS maintains several aggregators (e.g. Microgrids), named as Sub Service Provider (SSP) that manage customers/subscribers under their domains. The SSPs act as agents that perform local energy matching (inside their domains) and distributed energy matching within SSPs to determine the energy commitment. The goal of the distributed energy matching is to reduce the involvement of External Energy Supplier (e.g. \emph{Utility}) while providing a platform to demand side players to be a part of energy transaction. A distributed assignment problem is designed that requires minimum and aggregated information exchange (hence, secured) and solved by Linear Programming (LP) that provides the distributed matching decision. The communicative burden among SSPs due to the exchange of energy information is reduced by applying an adaptive coalition formation method. The simulations are conducted by implementing a synchronous distributed matching algorithm while showing the effectiveness of the proposed framework.
\end{abstract}


%
\IEEEpeerreviewmaketitle

\section{Introduction}
\label{intro}
As of 2014, the electricity market in Japan is dominated by regional monopolies, where 85\% of the installed generating capacity is produced by 10 privately owned companies. However, the rising of Power Producer and Supplier (PPS) (i.e. Electric Power Retailer) in the electricity market is inevitable due to the full-fledged deregulation \cite{meti} that will eventually break the monopolies. In this situation, PPSs can provide the platform to potential prosumers to bring their surplus of energy to the market and to actively participate in the energy transactions. The energy management within PPS is complicated and challenging with the integration of intermittent renewable energy sources. However, flexible power/energy operation as such is attainable through the advent of Digital-Grid architecture \cite{Abe:2011} and seminal concept like ``ECO net'' \cite{2003}.

The burden on \emph{Utility}\footnote{Hereafter, we refer any External Energy Supplier as \emph{Utility} for simplification} and the net energy balance within a PPS required to be minimized. Therefore, day-ahead optimal energy matching amongst the customers is essential. In this paper, we introduce a distributed energy commitment framework for the customers subscribed (subscribers) under PPS. Based on the energy supply and consumption pattern as well as the capability of participation on the electricity market, a subscriber commits; in a time-ahead fashion (e.g. day-ahead) to a certain energy profile for a particular time duration (e.g. 30-minutes) in future. Performing a centralized energy commitment considering the geographical locations and heterogeneity of subscribers is computationally costly, communicatively expensive and is exposed to Cyber-Physical vulnerability. Therefore, a distributed energy exchange scheme is proposed where the subscribers are grouped under several Sub-Service Providers (SSPs). The nature of SSP can potentially be that of a Microgrid or an Aggregator (can even be a virtual entity located in the vicinity of a PPS). The basic energy exchange problem is re-organized as 1) at lower level, energy exchange between consumers and producers, 2) at middle level, energy exchange between SSPs/aggregators/microgrids, and 3) at higher level, energy exchange with \emph{Utility}. As the size of PPS goes bigger, the communicative complexity at middle level increases. For this reason, a meshed network topology within SSPs is inefficient and practically infeasible. To solve this issue, a Learning based Coalition formation algorithm is adopted. This algorithm periodically provides the neighborhood map to the SSPs in order to reduce the communicative complexity for practical distribution system. The distributed matching scheme is essentially formulated as a distributed assignment problem where the subscribers are presented with their prior-engaged energy commitment, commitment capabilities coupled with their preferences over other subscribers.

We want to emphasize the fact that the system model presented in this paper is currently conceptual\footnote{The electricity market full-fledged deregulation is schedule to hit at 2020 in Japan.} with solid Business Models for the potential stake holders. For example, utilizing the proposed model, PPS makes profit by minimizing the energy transactions with \emph{Utility} in both spot and on-line imbalance market. On the other hand, the subscribers are paying less by purchasing energy locally (consumers) or earning more by selling energy locally (prosumers), thereby essentially increasing the share of renewable sources (so called customer-to-customer business model). Further motivation is drawn from the fact that the Feed-in-Tariff (FIT) scheme for encouraging higher renewable penetration will likely to be discontinued (or subjected to major reform, due to incompetence of \emph{Utility} to \emph{gridization} of renewables, especially PV), which makes prosumers open for merchandising the available renewables. The exact nature of the system model may vary with the PPSs emphasizing their own requirement, customer segments and overall Business Model.

\subsection{Related Works}
\label{rwork}
The Transactive Energy (TE) Framework \cite{TransactiveEnergy}(with follow-up standard-based architecture and protocols, e.g. \cite{TeMIX} and \cite{5759167}) follows the similar direction with an additional incorporation of different pricing schemes. The important distinction the proposed scheme brings is the influence of customer segmentation (detailed in later sections) and absence of detailed market-based control. The proposed method works with a minimum market-based control while depending on pre-engagement capability of the subscribers. However, due to its inherent distributed architecture, the proposed scheme can easily incorporate complex market based control. The distributed energy exchange problem can be realized through Multiagent framework where each SSP/Aggregator/Microgrid is represented as a potential agent. Works such as \cite{kok:08}, \cite{rad:10} are related to agent-based distributed operation and energy balancing. Given the multiagent framework with selfish agents (e.g. individual houses) in places, the agents can reach cooperative equilibrium through distributed optimization \cite{rad:10}. Other similar streams of important research regarding multiagent system (MAS) based energy management can be found in \cite{6179527}, \cite{6294477}, and \cite{6816377}. In \cite{6179527}, the authors designed two-level architecture of distributed resource management of multiple interconnected microgrids. In \cite{6294477}, the authors extended their work by incorporating demand response in the MAS-based resource management. We can align the proposed distributed energy matching framework with aforementioned (and similar types) researches. However, the important distinction between the proposed framework from the existing ones are the underlying physical power distribution system. The Digital Grid \cite{Abe:2011} architecture and ``ECO Net''\cite{2003} based power exchange lay the underlying physical power distribution assumption for the proposed framework. The existing distribution system (upon which most of the MAS based demand side EMS based on) is not as flexible as proposed in seminal papers \cite{Abe:2011} and \cite{2003}, where power can be exchanged within multiple customers at the same time. For example, a producer (e.g. generator in existing system, e.g. \cite{6179527}) can supply multiple consumers (e.g. loads) at the same time by tagging the power facilitating IP-based power tagging introduced in \cite{Abe:2011}. At the same time, a consumer also can receive energy from different producers. A small-scaled pilot program is already conducted \cite{dgdemo} (in Japan) that demonstrates the flexible power exchange within two residential units.

Moreover, the proposed framework takes the advantages of Microgrid coalition formation technique \cite{chakraborty:15} to periodically modify the communicative network topology. On the other hand, demand flexibility is considered to be essential entity of optimal demand side operation, especially with renewables. The research conducted by \cite{AlrOrtega2015705} and \cite{Evora20157456} provide a certification for DR program while realizing necessity for a standard interchange of DR and a direct load control for handling DR program, respectively. DR is becoming an integrated part of renewable-fueled future grid with energy storage. A stream of recent research e.g. \cite{HeydarianForushani2015941}, \cite{Gao2015275}, and \cite{6476049} tackle the issues related to robust scheduling of generation resources (that include renewables as well) considering energy storage with DR considering uncertainty in demand prediction and dynamic pricing. However, the interactions among demand side entities/players are not covered in these researches. Therefore, the distributed energy matching platform is a real necessity that operates in a deregulated electricity market with an extensive participation of Prosumers. In \cite{Izvercianu2014149}, the authors provide a value co-creation business model for Prosumer-oriented urban societies. Similar class of research regarding the role of Prosumers is also conducted by \cite{Zhang2015471} that essentially provides a game-theoretic operational scheme for residential distribution system with high number of participation from prosumers. The present research advances the state-of-the-art of prosumer-centric deregulated energy society by creating a energy commitment based platform that tries bind together all emerging energy players those otherwise are unable to participate in electricity market.
\begin{figure}[H]
\centering
\includegraphics[scale=0.2]{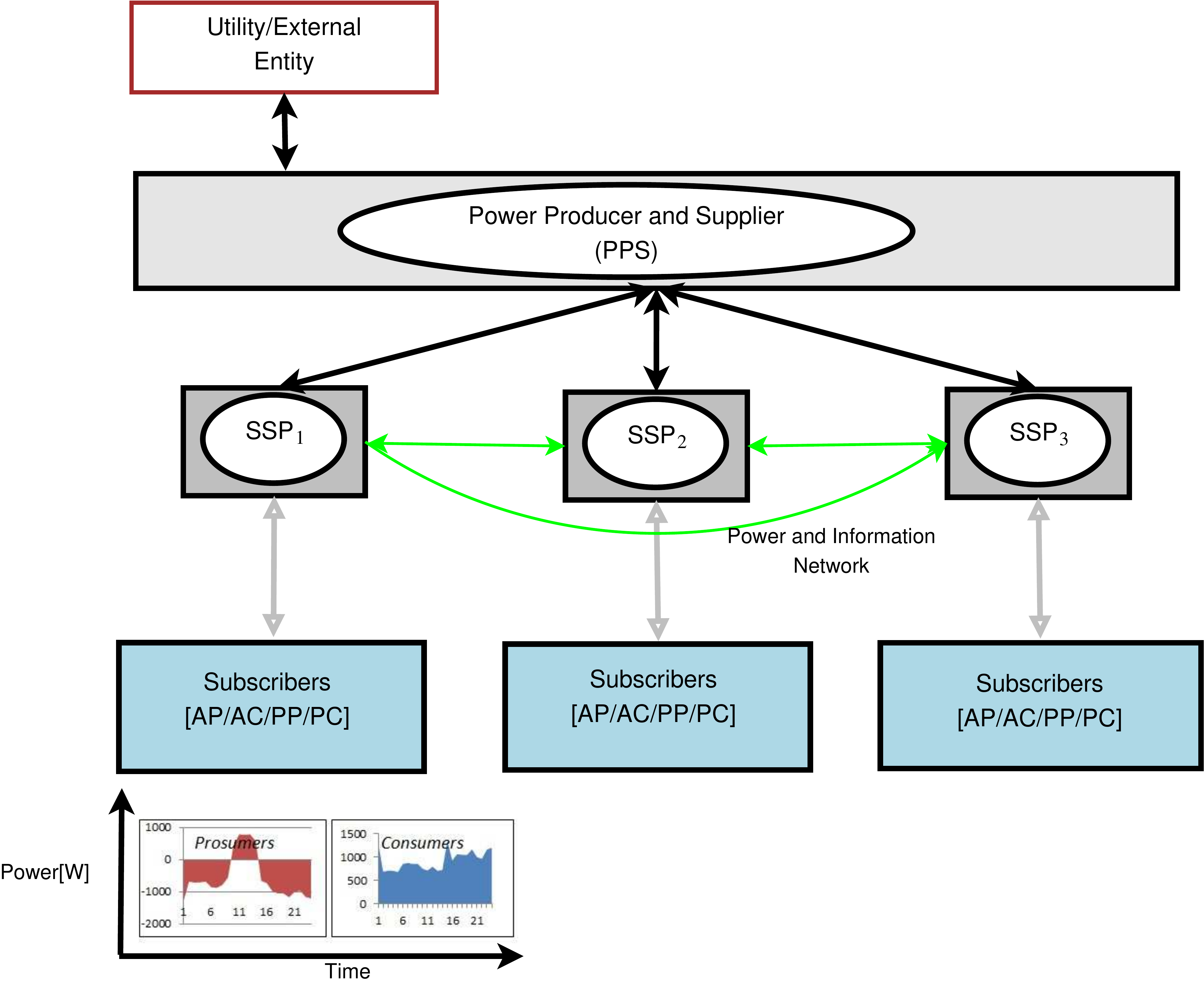}
\caption{System outline for Distributed Energy Commitment (for 3 SSPs)}
\label{dist_model}
\end{figure}

\section{System Architecture}
\label{sysModel}
The basic architecture of the proposed scheme and model is outlined in Figure~\ref{dist_model}. The PPS manages energy exchange among several SSPs operating under that particular PPS. The number of SSPs depends on the service region, service specification and geographical coverage of that PPS. The PPS is thus responsible for dealing and managing energy and power within the specific service region. The breakdowns of subscribes are provided below.

\subsection{Subscribers' Breakdown}
\label{subs}
The subscribers are primarily either producers or consumers or a combination of both (so called prosumers). A further breakdown to each subscriber group is possible considering the energy profile and commitment of the subscribers to the service. The commitment in this context is defined as the willingness (with capability) to consume or produce certain amount of energy (that is committed earlier) for the next day (or another period in future). The brief description of each group of subscribers is presented below.

\subsubsection{Active Producer}
\label{AP}
Active Producer (AP) is a special class of energy producer that is committed to participate in distributed energy exchange scheme. 
Such commitments from producers are integrated into the system that leads to a better and efficient energy management. An AP is able to declare the energy production before a certain period of time. Typically, it can be realized as a day-ahead based scheme. 

\subsubsection{Passive Producer}
\label{PP}
Passive Producer (PP) is a special class of AP that can provide flexibility over the energy production. The SSP (or PPS) utilizes the flexibility feature provided by the PP if necessary. The PP can use their Spinning and Operating Reserve capacity with ramping ability to incorporate such signal. The SSP however, doesn't utilize the whole flexibility in day-ahead operation. Rather, the SSP keeps a certain fraction of the declared flexibility for mitigating on-line deviation. Moreover, PV/Wind based DERs are not able to declare their energy production precisely for a future period due to the uncertainty in forecasting. In such case, that kind of producers will fall into the category of PP and will declare the estimated production amount with associated confidence (reported as flexibility).

\subsubsection{Active Consumer}
\label{AC}
Active Consumer (AC) is classified as the consumer who joins the scheme and provides the future energy 
demand (in the form of prior-engaged capability) and a list of preferred APs
from which that AC wants to receive energy. The preference list gives some control to AC over choosing their preferred energy break down. For example, an AC may prefer renewable powered APs over other APs. 

\subsubsection{Passive Consumer}
\label{PC}
A Passive Consumer (PC) is a special class of AC that does not commit entirely regarding the energy consumption and also should be ready to compromise on the energy usage. For example, in a DR program (as a direct control), the SSP issues a signal to a particular PC to shut-down one or more power-hungry devices. Based on the flexibility (reported as a percentile of demand reduction), the PC can react to such signal.

\subsection{Interactions Among Players}
\begin{figure}
\centering
\includegraphics[scale=0.42]{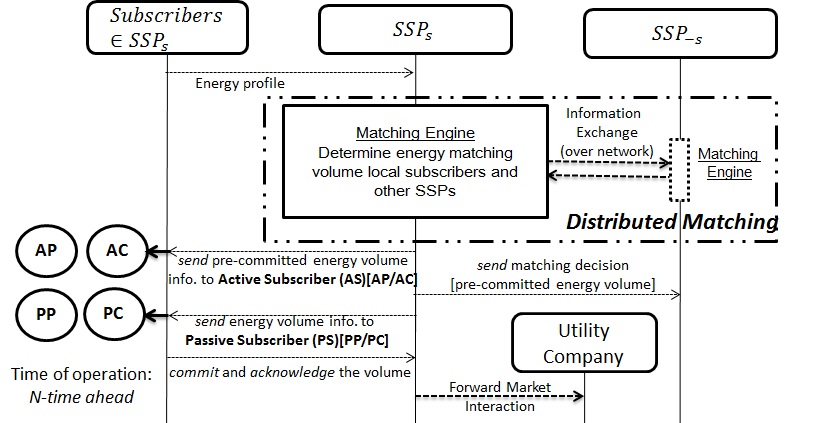}
\caption{Interactions and time sequence within different players}
\label{time_interaction}
\end{figure}

Figure~\ref{time_interaction} shows the high-level interactions among subscribes (operating under $SSP_s$), \emph{Utility} and other SSPs (except $s$, $SSP_{-s}$). The whole operation is divided into two phases. Phase 1 is operating on an \emph{N}-time ahead (e.g. day-ahead) level while Phase 2 is operating on real-time level. In Phase 1, the subscribers are assumed to provide their predicted energy profiles for a certain future period. 
The distributed energy matching operation within local subscriber and external SSPs is performed in this phase. 
The distributed energy matching operation is actually solving a distributed optimization problem which tries to attain local matching objectives while taking the external SSPs energy status into account in order to achieve the objectives of global matching, energy balancing and reduced interaction with \emph{Utility}. 
The matching operation takes care of different preferences provided by the subscribers (Active and Passive) and provides them the decision regarding the volume of energy needed to be exchanged (in a future period). Such decisions come as commitment for both active and passive subscribers. The matching engine also provides the unavoidable (yet required) energy interactions to the \emph{Utility} as a forward market interaction. The associated SSPs (also based on the decision of distributed matching) are also informed with the required energy transactions. In real-time operation, the actual energy exchange (based on the committed volume of energy) takes place within the subscribers and \emph{Utility}. However, due to the external factors, sudden change in energy demand and energy supply might occur which creates deviation between supplied energy and total demand. The detailed real-time operation is exempted in this article since it requires elaborate descriptions.

\subsection{Workaround Examples}
This section presents an examples to clarify the framework and energy matching process.

\begin{figure}
\centering
\includegraphics[scale=0.33]{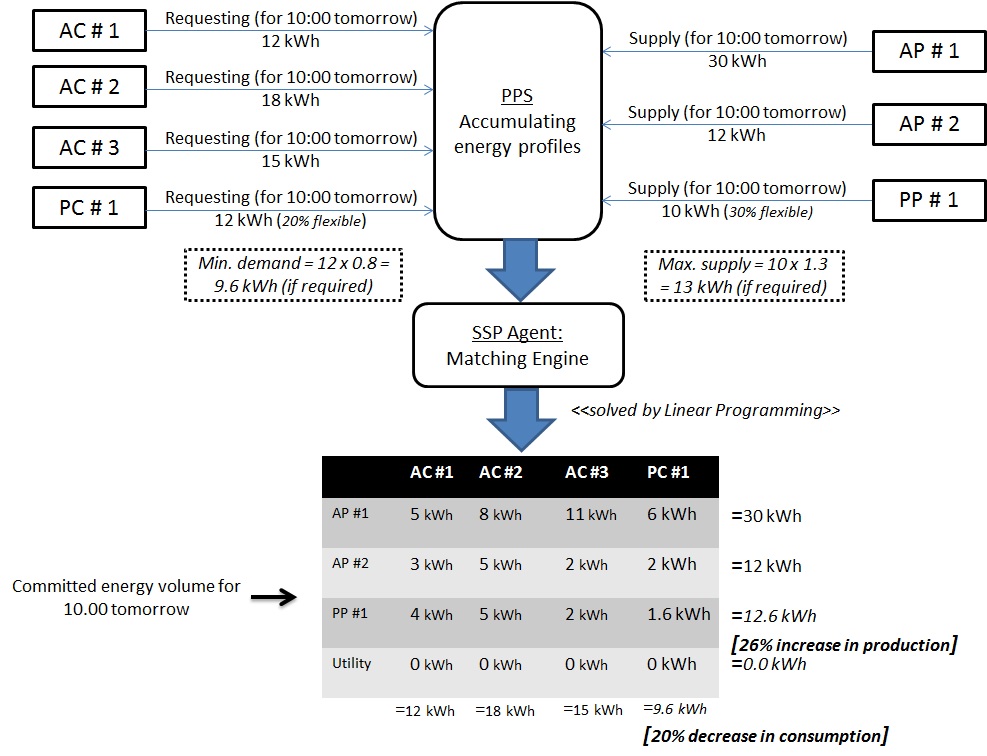}
\caption{Example of energy matching within multiple active and passive subscribers}
\label{example_e_matching}
\end{figure}

\subsubsection{Energy Matching}
\label{energy_matching}
Figure~\ref{example_e_matching} shows a simple exemplary energy matching procedure among 3 Active Consumers (AC), 1 Passive Consumer (PC), 2 Active Producers (AP), and 1 Passive Producer (PP) for a day-ahead operation (accord to N-time ahead operation). 
The energy matching will be conducted for 10 AM in next day. This example assumes that, the consumers and producers are able to provide their own energy profiling. 
Therefore, the prediction engine is not in the action for this example. Passive subscribers (PC and PP) come up with their flexibility of 20\% and 30\%, respectively. 
In the case of the PC, 20\% flexibility refers to the reduction of demand down to 20\% (i.e. for PC \#1, it can bring down the demand from 12 kWh to 9.6 kWh, if the PPS instructs to do so). And, in case of the PP, 30\% flexibility refers to the increase in production up to 30\% (i.e. for PP \#1, it can increase the production from 10 kWh to 13 kWh, if the PPS instructs to do so). After the PPS accumulates all the requested and potential supply quantity of energy from consumers and producers, respectively, the energy matching operation started. Finally, the output is tabled summarizing the energy transactions among consumers and producers. It is noted that, the total supply is 52 kWh to 55 kWh (with the flexibility of PP \#1) and the total demand is 57 kWh (can be reduced to 54.6 kWh with the flexibility of PC \#1). In the ideal case with no flexibility (i.e. demand is 57 kWh and supply is 52 kWh), the \emph{Utility} will be required to provide additional 5 kWh to nullify the gap between supply and demand. 

However, the PPS utilized the flexibility of passive customers and zeros the \emph{Utility} interaction by instructing PC \#1 to reduce the consumption down to 20\% and PP \#1 to increase the production up to 26\% (out of 30\% flexibility). The resultant table can be read as (e.g. row 1); AP \#1 is committed to provide 5, 8, 11 and 6 kWh to AC \#1, AC \#2, AC \#3, and PC \#1, respectively at 10 AM tomorrow. The same way, column 2 can be read as; AC \#2 is committed to receive 8, 5, 5 kWh of energy from AP \#1, AP \#2 and PP \#1, respectively and no energy from the \emph{Utility}. The decision on energy matching can have multiple optima (i.e. multiple solutions can be achievable while realizing the same objective). 
However, depending on criteria such as, the preferences (such preference may be, e.g. AC \#1 prefers AP \#1 over AP \#2 to provide higher fraction of requested energy), fairness policy (such policy may be, e.g. PPS provides certain advantage to AC \#1 while respecting AC \#1’s preference), etc., a single solution can be attainable. These features can be included while design specific services. In the real time operation, however, the PPS might ask passive customers (that contain the flexibility unused in the day-ahead operation) to adjust the real time demand-supply gap. For instance, in the presented example, PP \#1 can increase the production slightly (4\%) in the real time operation (possibly by utilizing the operating reserve).

\subsection{Pricing Assumption}
Pricing assumption is very important while economically modeling and realizing the system. The designed pricing mechanism should be able to provide enough incentives to all the stakeholders to join the scheme. The main motivations of pricing design are
\begin{itemize}
\item[1.] A PP/AP should sell per unit energy to PPS in a price higher than what that PP/AP sells per unit energy to \emph{Utility}
\item[2.] A PC/AC should buy per unit energy from PPS in a price lower than what that PC/AC uses to buy from external \emph{Utility}
\end{itemize}
These assumptions are ensured by the PPS. Moreover the subscribers are assumed to provide their true predictions regarding energy consumption or production. The subscribers, therefore, are in a \emph{No-Game} situation and avoid strategic interactions. Strategic interactions are, however, obvious when SSPs are exposed to different pricing environment (stronger market-based control). For example, a PC that participates in DR program evidently provided with economic incentive to act as one. On the other hand, a PP should receive additional monetary value in case of activating the spinning reserve. At the same time, the PPS has to ensure that a subscriber does not deviate significantly from its commitment by implementing pricing scheme as \cite{Chakraborty:2014}. In this case, \emph{Game Theory} based analysis can be applied to \emph{Nash}-out the situation in order to find the associated equilibrium. An immediate follow-up research will concentrate on appropriate pricing mechanism design for subscribes as we limit this contribution to framework introduction and required matching algorithm.

\section{Distributed Matching Operation}
\label{dist_architecture}
An SSP realizes the following objectives through a distributed optimization problem, i) minimizing the energy transactions with \emph{Utility}, ii) maximizing the local energy transactions within local customers, iii) respecting the preference of consumers (A/P)C. 

\subsection{Objective Function with Constraints}
The multiple objectives for the above optimization problem are described in this section. A particular SSP, $s$ tries to
attain the multi-objectives, defined in (\ref{main_obj_eq}). The set of SSPs working under the particular PPS other than $SSP_s$ is assumed as $SSP_{-s}$. For notational simplicity, (A/P)C is presented by $AC$ and (A/P)P is represented by $AP$. Eq, (\ref{main_obj_eq}) is the scalarized multi-objective function with weights corresponding individual objective.

\scriptsize
\begin{equation}
\overset{min}{\substack{cm(i,j),\\fx(i),fx(j)}}\left\{\begin{matrix}
-w_{1}\times&\left [ \displaystyle \sum_{i\in AC_s\setminus \{U\}}{}Pr(i) \sum_{j \in AP_s\setminus \{U\}}{}cm(i,j) \right ]\\ 
+w_{2}\times&\left [ \displaystyle \sum_{i\in AC_s}cm(i,U) \right ]\\ 
-w_{3}\times&\left [ \displaystyle \sum_{i\in AC_s\setminus \{U\}}\sum_{j \in AP_s \setminus \{U\}}{}(cm(i,j)+\alpha\times[\beta-pt(i,j)]) \right ]\\ 
-w_{4}\times&\left [ \displaystyle \sum_{i\in AC_s\setminus \{U\}}\sum_{j \in SSP_{-s}}{}cm(i,j) \right ]\\ 
-w_{5}\times&\left [ \displaystyle \sum_{i\in AC_s\setminus \{U\}}\sum_{j \in SSP_{-s}}{}(cm(i,j)+\alpha\times[\beta-pt(i,j)]) \right ]\\ 
\end{matrix}\right.
\label{main_obj_eq}
\end{equation}
\normalsize

In (\ref{main_obj_eq}), $w_{1}$ presents the weight related to regulatory objective function. This objective function ensures the Active subscribers are served before Passive ones, which is denoted $Pr(i)$. $Pr(i)$ is typically, decided by the contract between PPS and subscriber $i$. The \emph{Utility} is exempted from the list of ACs. The objective function weighted by $w_{2}$ minimizes the energy exchange with \emph{Utility}. The preference of each subscriber is respected by the objective function associated with the weight $w_{3}$. The energy information received from other SSPs (described by $SSP_{-s}$) are handled in the objective functions weighted by $w_4$ and $w_5$. The weights $w_4$ and $w_5$ describes maximizing the energy exchange with other SSPs and maximizing the preferences of ACs (belonging to $SSP_s$) towards other SSPs, respectively. While analyzing (\ref{main_obj_eq}) it can be noticed that, the objectives weighted by $w_{1}$ and $w_{4}$ are essentially carrying the same variables. In objective weighted by $w_{1}$, the index $j$ represents a member of the local producers set, $AP_{s}$. While in objective weighted by $w_{4}$, the index $j$ is the set of other $SSP_{-s}$ that are physically connected with $SSP_{s}$. Therefore, we can easily merge these two objectives into one. Similarly, objectives weighted by $w_{3}$ and $w_{5}$ can be merged into one. By combining these similar objectives in (\ref{main_obj_eq}), (\ref{mod_obj_eq}) is formed as follows.

\begin{equation}
\overset{min}{\substack{cm(i,j),\\fx(i),fx(j)}}\left\{\begin{matrix}
-w_{14}\times&\left [ \displaystyle \sum_{i\in AC_s\setminus \{U\}}{}Pr(i) \sum_{\substack{j \in AP_s\cup SSP_{-s}\\ \setminus \{U\}}}{}cm(i,j) \right ]\\ 
+w_{2}\times&\left [ \displaystyle \sum_{i\in AC_s}cm(i,U) \right ]\\ 
-w_{35}\times&\left [ \displaystyle \sum_{i\in AC_s\setminus \{U\}}\sum_{\substack{j \in AP_s\cup SSP_{-s}\\ \setminus \{U\}}}{}[\substack{(cm(i,j)+\\\alpha\times[\beta-pt(i,j)])}] \right ]\\ 
\end{matrix}\right .
\label{mod_obj_eq}
\end{equation}

Where $U$ represents the \emph{Utility}. The decision variables $cm(i,j)$ and $fx(i)$ or $fx(j)$ are the committed energy to be exchanged between subscriber $i$ and $j$, and flexibility of commitment, respectively. Since, (\ref{mod_obj_eq}) is an objective function for distributed matching operation, at certain times $j \in AP_s \cup SSP_{-s} \setminus \{U\}$ represents the set of local producers and different SSPs (other than $SSP_{s}$) except the \emph{Utility}. The weight vector $w$ is calibrated utilizing a \emph{local search} algorithm. Come to the description of (\ref{mod_obj_eq}), $Pr(i)$ represents the serving priority of (A/P)Cs which will be respected by the SSP while deciding. Since local energy transaction is preferred over the transaction with other SSP (which in turn, preferred over the transaction with \emph{Utility}), a priority table (as $pt(i,j)$ is defined (in the 3rd line of (\ref{mod_obj_eq})) that describes the preference. 
Basically, the transaction order can be defined as the following preference relation ($inSSP$ is transaction inside SSP, and $outSSP$ is the transaction outside SSP but inside PPS with other SSPs).

\begin{equation}
inSSP\succ outSSP \succ Utility
\label{pref}
\end{equation} 

The $pt(i,j)$, in case of local transaction, defines the preference of customer $i$ regarding energy source (e.g. if $i$ prefers green energy over cheap energy, it will prefer a certain $j$ over others). Back to (\ref{mod_obj_eq}), the 2nd part objectifies the minimization of energy exchange with \emph{Utility}. Note that, for 
the transactions defined in 1st and 3rd part, no \emph{Utility} is involved. The above objective function is subjected to the following constraints ($\forall i \in \{AC_s \cup\{U\}\}$ and $\forall j \in \{AP_s \cup SSP_{-s} \cup \{U\}\}$) 

\begin{equation}
\sum_{i\in AC_s}cm(i,j) \leq fx(j) Ep(j)
\label{const_producer}
\end{equation} 

\begin{equation}
fx(i)Dc(i) \leq \sum_{j\in AP_s}N(i,j)\times cm(i,j) \leq Dc(i)
\label{const_consumer}
\end{equation} 

\begin{equation}
[1-bound(i)] \leq fx(i)\leq 1, \forall i \in AC_s
\label{flex_bound_ac}
\end{equation} 

\begin{equation}
1 \leq fx(j) \leq [1+bound(j)], \forall j \in AP_s
\label{flex_bound_ap}
\end{equation} 

We assume, the \emph{Utility}, $U$, can sell or buy any amount of energy. (\ref{const_producer}) constraints the total supply should not exceed the total production ($Ep(j)$) of a producer. (\ref{const_consumer}) constraints the total demand, i.e. $Dc(i)$ of a consumer must be met. The terms $fx(i)$ and $fx(j)$, constrained by (\ref{flex_bound_ac}) and (\ref{flex_bound_ap}), respectively define the associated flexibilities. In case of Active Subscriber (i.e. AP or AC), $fx=1$ (by placing $bound=0$). The $bound(k)$ defines the maximum flexibility a passive subscriber can afford. For example, if a PC can reduce 20\% of energy consumption, its $bound$ will be \emph{0.2}. Similarly, if a PP can increase 10\% of its committed production, its $bounds$ will be \emph{0.1}. Therefore, the optimizer can decide $fx$ within \emph{0.8} to \emph{1} for a PC and \emph{1} to \emph{1.1} for a PP. The transmission line constraints can be implemented by bounding the commitment matrix $cm$ within the upper and lower power flow capacity. Considering the loss function $\mathcal{L}$ with lower and upper power flow capacity, $\Gamma_{min}$ and $\Gamma_{max}$ (respectively), the simple transmission line constraint can be implemented as (\ref{eq_tlc_bound}). The reformed distributed matching problem equivalently casted as Security Constrained Unit Commitment (SCUC) and can be efficiently solved utilizing methods such as \cite{6680791}. In this paper, however, the SCUC is not considered.

\begin{equation}
\Gamma_{min}(i,j) \leq \mathcal{L}(cm(i,j)) \leq \Gamma_{max}(i,j), \forall i,j \in {SSP \bigcup {U}}
\label{eq_tlc_bound}
\end{equation}

\begin{figure}[h]
\centering
\includegraphics[scale=0.33]{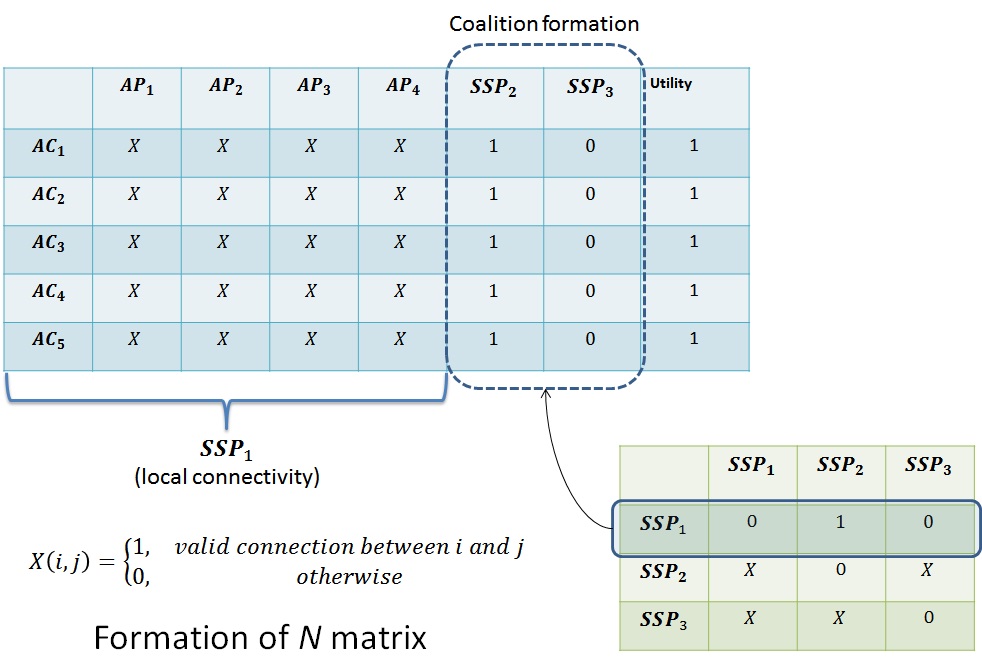}
\caption{Exemplary formation of binary $N$ matrix}
\label{matrix_N}
\end{figure}
The binary matrix $N$, in (\ref{const_consumer}) presents the local physical connectivity between two subscribers. The matrix $N$ is provided by the PPS and is periodically updated reflecting the demand and supply profile. The mapping $N$ is envisioned to contain both local and distributed communication and physical network infrastructure; the local portion of the mapping contains connectivity information within $SSP_{s}$ where the distributed portion represents the connectivity information (that is an outcome of a \emph{learning based coalition formation} algorithm, detailed in Section \ref{comm}) among $SSP_{s}$ and $SSP_{-s}$. The formation of $N$ matrix is shown in Figure~\ref{matrix_N} for an exemplary case of 3 SSPs. The tables in figure are drawn from the perspective of $SSP_{1}$ comprising of 5 ACs and 4 APs. The inter SSP connectivity is determined by the \emph{learning based coalition formation} engine. SSPs exchanges the aggregated energy (surplus information that is essentially the outcome of the distributed optimization) that makes them potential producers in the distributed matching process. In this example, $SSP_{1}$ can exchange energy with $SSP_{2}$, but not with $SSP_{3}$. The objective function associated with weight $w_{4}$ in (\ref{mod_obj_eq}) shows the aggregated decision variable $cm(i,j)$ of SSP, $j$ ($\in SSP{-s}$) with local consumers $i$ ($\in AC_{s}$). The distributed matching operation thus requires aggregated energy surplus information from a neighboring SSP, $j$, as appeared in the constraint (\ref{const_producer}) by $Ep(j)$. The SSP, $j$ also needs to provide flexibility bound for $fx(j)$ for the constraint in (\ref{flex_bound_ap}). The $bound(j)$ for an SSP, $j$ is calculated as (private calculation of $SSP_{j}$ that is hidden to other SSPs)

\begin{equation}
bound(j) = \frac{\sum_{l\in SAP_{j}}\left [ \left \{ 1+bound(l) \right \} \times EP(l) -cm(.,l) \right ]}{\sum_{l\in SAP_{j}}[Ep(l)-cm(.,l)]} - 1
\label{bound_j}
\end{equation}

where, $SAP_{j}$ is the set of $AP$s in $SSP_{j}$ that are able to supply energy, i.e. $[Ep(l)-cm(.,l)]>0$ ($\forall l \in AP_{j}$) and $cm(.,l)$ is the total commitment of energy production for $l$ to the local ACs of $SSP_{j}$. The associated flexibility bound is provided by $bound(l)$. In other words, (\ref{bound_j}) provides the weighted flexibility bound calculated over all valid producers who are able to provide energy.

\subsection{Privacy Measure}
\label{sec_measure}
The designed distributed optimization formulation in (\ref{mod_obj_eq}) has a special feature that counts into the privacy of each subscriber. Certain regulations (e.g. in Japan and EU countries) do not allow the subscribers (or consumers) to share their private energy information with peers (e.g. neighbor subscribers) due to the security reason (i.e. sharing smart-meter data)\footnote{Although the associated policy is still under discussion, an NIST guideline (NISTIR 7628 \cite{nistir:7628}) regarding Privacy is likely to be followed in Japan as well.}. Therefore, most of the multiagent based distributed optimizations (e.g. \cite{rad:10}) that require exchanging energy information with each other in order to reach a collaborative optimized energy profile (e.g. peak reduction, demand response, etc.) become obsolete in situation as such. The aggregated energy profile of subscribers, however, does not have such issue since it hides the exact energy information of individual subscriber (that is private to that particular subscriber). In the proposed architecture, only a designated controller (i.e. the SSP, where the distributed optimization algorithm is hosted) has the exact energy breakdown information of each of its subscribers. No subscriber knows its peer's energy profile while partaking in the distributed optimization. Therefore, local privacy measures of consumers are sustained that makes the proposed distributed energy matching scheme applicable in a relatively private society. Moreover, while performing the distributed energy matching with neighboring SSPs (and with the \emph{Utility}), a particular SSP only shares its aggregated energy surplus information coupled with associated aggregated flexibility instead of the entire energy information of its subscribers. The formation of distributed optimization in (\ref{mod_obj_eq}), with constraint as (\ref{const_producer}), ensures that only aggregated energy profile is required to reach the convergence (i.e. $Ep(j);\forall j\in SSP_{-s}$).

\subsection{Distributed matching algorithm}
The objective function and constraints are linear in nature that makes the distributed assignment problem solvable by LP. The weights $w_{14}$, $w_{2}$, and $w_{35}$ describe associated preferences for each of the objective functions and can be fixed by a PPS. The detailed algorithm is shown in Algorithm \ref{DM}. The Algorithm \ref{DM} describes the operation of a particular SSP while performing the local and distributed optimization. The algorithm takes into the demand and energy capacity of consumers and producers, respectively with associated flexibility limits. Additionally, it also requires the preference of local consumers (over local producers) and fairness policy of local consumers. The outcomes of Algorithm \ref{DM} are the optimal energy commitment within local subscribers as well as peer SSPs (that are physically connected and are belonging to same coalition) and the optimal flexibility of passive subscribers. An interesting feature of Algorithm~\ref{DM} is that, the broadcasted (energy) surplus information is implemented using a synchronous lock. Doing that ensures that a particular load SSP takes the maximum available surplus energy (from a generator SSP) to balance its load. Therefore, the energy commitment matrix $cm$ is already integrated the minimum possible energy exchange (Figure~\ref{graph_matching} clears the point).

\begin{algorithm}
\DontPrintSemicolon
\KwData{$Ep(j)$ energy capacity of all local producers, $j$ in $SSP_{s}$ and other SSPs, $j$ in $SSP_{-s}$.}
\KwData{$Dc(i)$ demand of all local consumers, $i$ in $SSP_{s}$.}
\KwData{$bound(j)$ flexibility bound of all subscribers.}
\KwData{$Pr(i)$ fairness priority of local consumers such that $\sum Pr(i) = 1$.} 
\KwData{$pt(i,j)$ preference of a consumer $i$ over a producer $j$.}
\KwData{$N(i,j)$ connectivity matrix between consumer/SSP $i$ and producer/SSP $j$.}

\KwResult{$cm(i,j)$ the energy commitment from producer $j$, $\forall j \in \{AP_s \cup SSP_{-s} \cup \{U\}\}$ to consumer $i$, $\forall i \in \{AC_s \cup \{U\}\}$.}
\KwResult{$fx(j)$ $\forall j\in \{AP_{s} \cup AC_{s}\}$ flexibility bound of all subscribers.}
\Begin{
$bestSolution \longleftarrow \infty$\;
\For{$k\in SSP_{-s}$}{
$Ep(k) \longleftarrow 0$\;
}

$statusChanged \longleftarrow True$\;
\While{$statusChanged$}{
$statusChanged \longleftarrow False$\;
$tCM, tFX, currentSolution$ $\longleftarrow$ solveDistMatching($Ep$, $Dc$, $bound$, $Pr$, $pt$, $N$)\; 

\If{$currentSolution < bestSolution$}{
$bestSolution \longleftarrow currentSolution$\;
$statusChanged \longleftarrow True$\;

$exEnergy \longleftarrow 0$\;
$totalEnergy\longleftarrow 0$\;
\For{$j \in AC_{s}$}{
\If{$[(1+bound(j))\times Ep(j) - cm(.,j)] > 0$}{
$exEnergy \longleftarrow exEnergy + [(1+bound(j))\times Ep(j) - cm(.,j)]$ 
$totalEnergy \longleftarrow exEnergy + [Ep(j) - cm(.,j)]$\;
}
}

$tBound \longleftarrow \frac{exEnergy}{totalEnergy+1\times10^{-7}} - 1$\;
\If{$exEnergy > 0$}{
\For{$SSP_{k} \in \{ l \in \text{shuffle}(SSP_{-s}) | N(s,l)~is~True\} $}{
$broadCastEnergy(SSP_{k}, SSP_{s},$\;
$exEnergy, tBound, SEND\_EXCESS)$
$exEnergy \longleftarrow exEnergy - cm(SSP_{k},SSP_{l})$ \;
}
}
$cm \longleftarrow tCM$ \;
$fx \longleftarrow tFX$ \;
}
$SSP_{dst}, SSP_{src}, supply, sBound, token = blockMessageReceive()$ \;
\If{$supply > 0~\text{\bf and}~SSP_{s}==SSP_{dst}~\text{\bf and}~token==SEND\_EXCESS$}{
$statusChanged \longleftarrow True$\;
$Ep(SSP_{src}) \longleftarrow supply$\;
$bound(SSP_{src})\longleftarrow sBound$\;
}
}
}
\caption{DistributedMatchingForSSP($SSP{s}$)\label{DM}}
\end{algorithm}
\begin{figure}[H]
\centering
\includegraphics[scale=0.25]{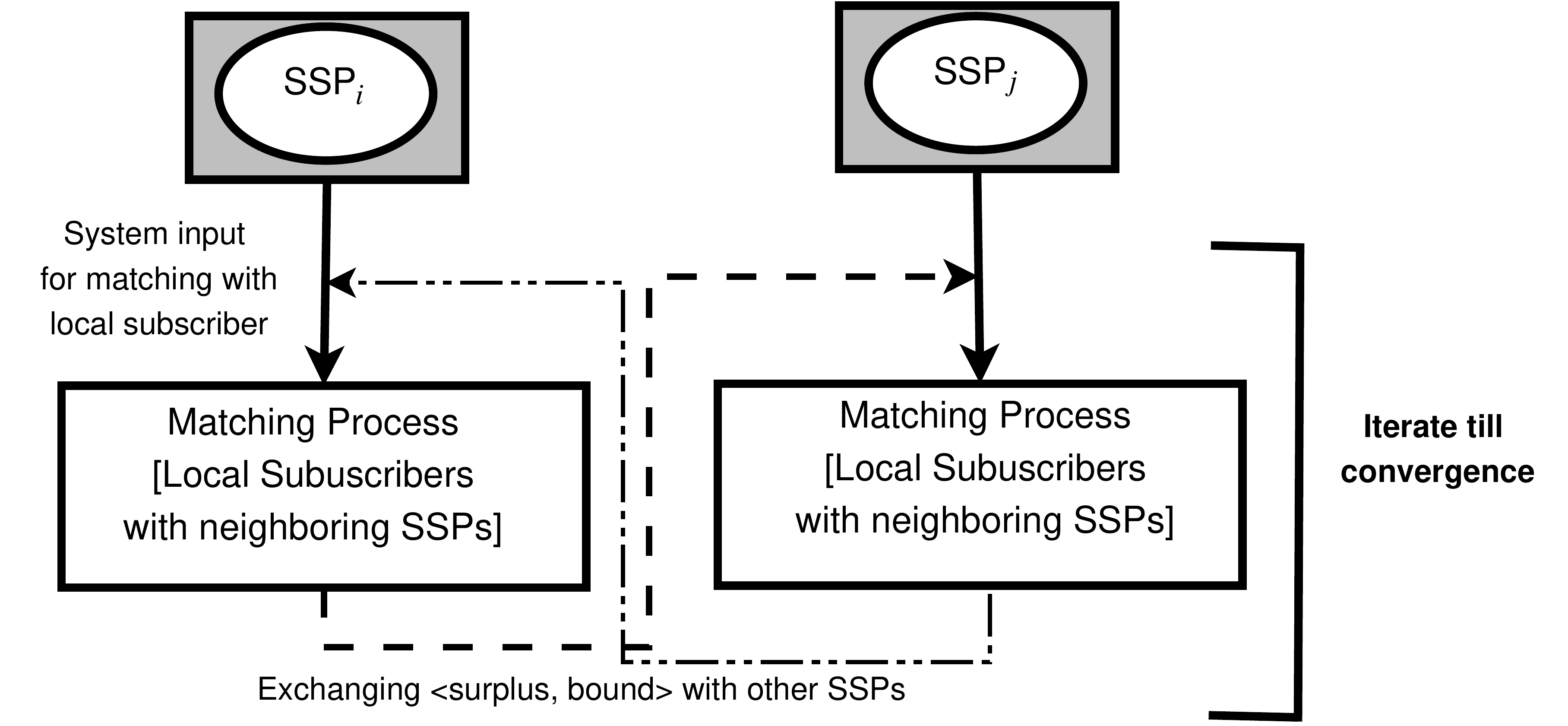}
\caption{Distributed energy matching process between two SSPs}
\label{dist_e_matching}
\end{figure}

The distributed matching process between two SSPs is shown in Figure~\ref{dist_e_matching}. For instance, $SSP_{i}$ and $SSP_{j}$ iteratively optimize their local energy exchange coupled with energy exchange within themselves. These two SSPs continue to do such distributed exchanging until a certain convergence criterion is encountered. The distributed matching will converge when no surplus of energy is available.



\begin{figure}[h]
\centering
\includegraphics[scale=0.43]{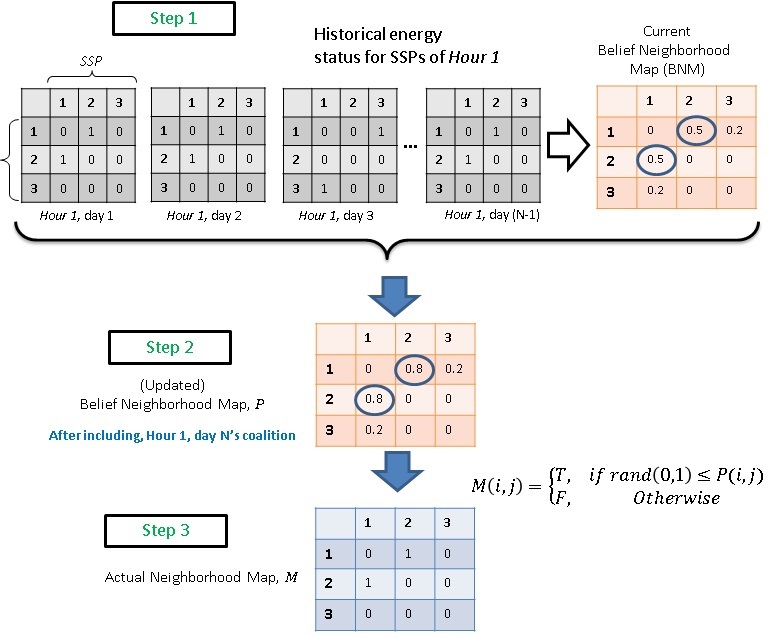}
\caption{Network map creation example}
\label{network_map_example}
\end{figure}

\subsection{Network Mapping by Learning based Coalition Formation}
\label{comm}
The interconnections of the communicative and power network within the SSPs tend to increase exponentially with the increase in number of SSPs. Therefore, it is essential to reduce the network complexity without much compromising with optimality. 
To remedy this issue, a neighborhood map generation process is designed. The process essentially takes the advantage of Microgrid Coalition Formation method \cite{chakraborty:15} with historical coalition formation information thereby, periodically updates the neighborhood map of each SSP. The coalition formation engine utilizes the periodically energy status of each SSP (i.e. surplus or deficit of energy information) to create groups of interconnected SSPs. The groups (and the SSPs in them) are optimally created so that the energy interactions inside a group are maximized. 

The PPS utilizes the historical energy status of each SSP where the Coalition Engine in PPS determines the energy based coalition among SSPs (step 1 (S1)). The Coalition Engine maintains a Belief Neighborhood Map (BNM) that contains a skeleton of Neighborhood Map using a Probabilistic Prior (step 2, S2). The BNM is created by a \emph{Bayesian} approach that takes into account the periodical formed coalitions. An exemplary creation of BNM is shown in Figure~\ref{network_map_example} considering three SSPs (1, 2, and 3) for a particular period (Hour 1). As seen from the figure, the current BNM (which was created by statically analyzing the Hour 1's energy status of past N-1 days) is updated when the N-th day's energy status of Hour 1.
An Actual Neighborhood Map (ANM) is generated by taking a snapshot of BNM. The snapshot is essentially probabilistic realization of BNM (shown in Figure~\ref{network_map_example}. The PPS delegates the ANM only if there is a significant update in \emph{updated BNM} (in Step 3) compared to \emph{current BNM}. As seen in Figure~\ref{network_map_example}, the probability that SSP\#1 and SSP\#2 are in same coalition changes from \emph{0.5} to \emph{0.8} after incorporating the new evidence. In this case, therefore, the PPS delegates the ANM to associated SSPs. The distributed matching operation (in SSP) utilizes the ANM (or an updated ANM in case of delegation from PPS) to perform matching operation with the exchange of energy information with other SSPs. The ANM is the (portion of) binary network matrix $N$ that was defined in Algorithm \ref{DM}. In step 4, the off-line matching decisions are provided as shown in Figure~\ref{time_interaction}. The updated energy status (prior to matching operation) of that particular SSP is sent back to PPS's Coalition Engine (in step 5) to update the BNM.

\section{Experiments and Analysis}
\label{experiment}
This section presents some experiments with associated discussions that are conducted to show the effectiveness of the proposed distributed matching scheme. Tokyo's residential demand data (with/without PV installation) are taken and scaled up by adding random variance (typically, \emph{Normal Distribution}). Several case studies are presented.

\begin{figure}[h]
\centering
\includegraphics[scale=0.35]{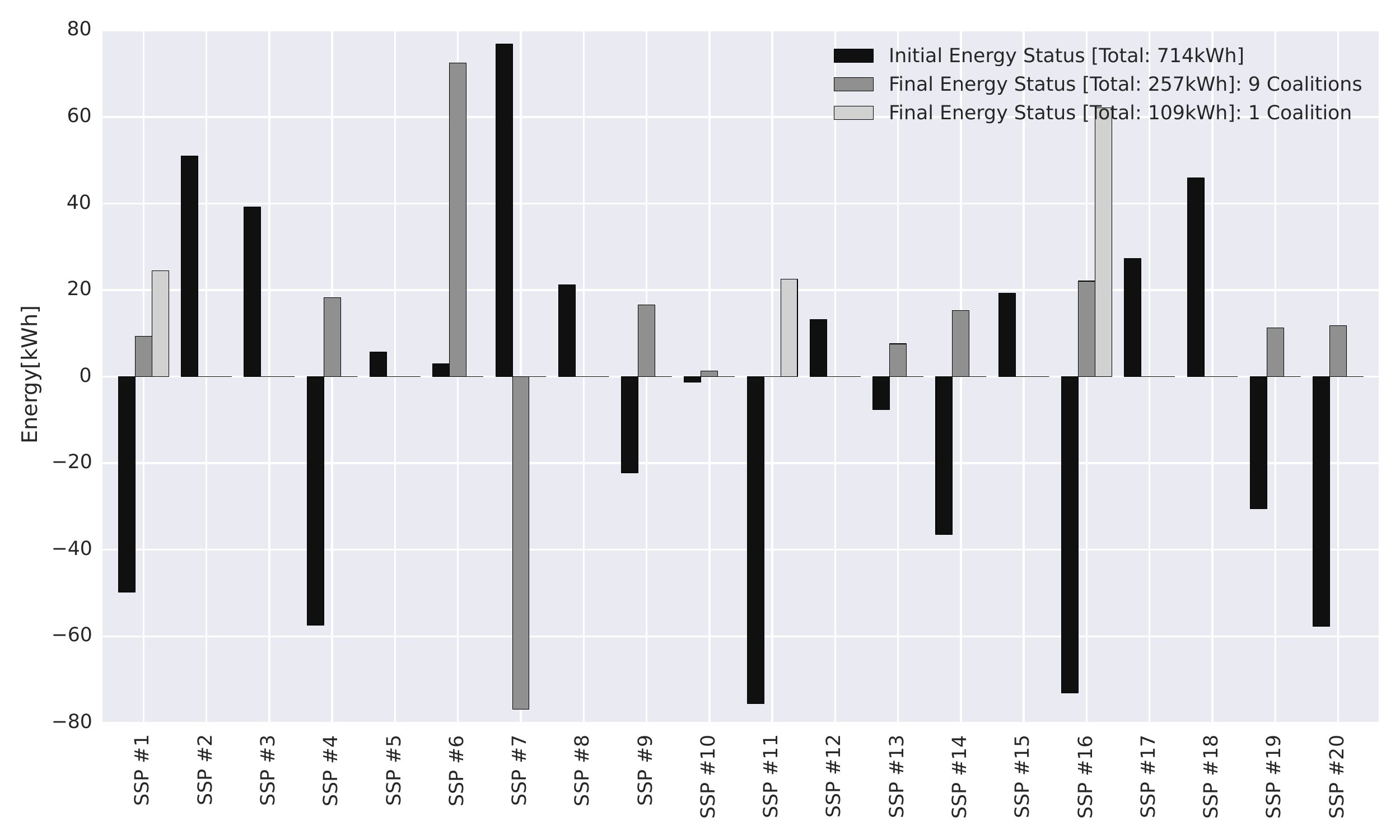}
\caption{Comparisons of SSP-wise absolute energy status}
\label{graph_e_status}
\end{figure}

\begin{figure}[h]
\centering
\includegraphics[scale=0.6]{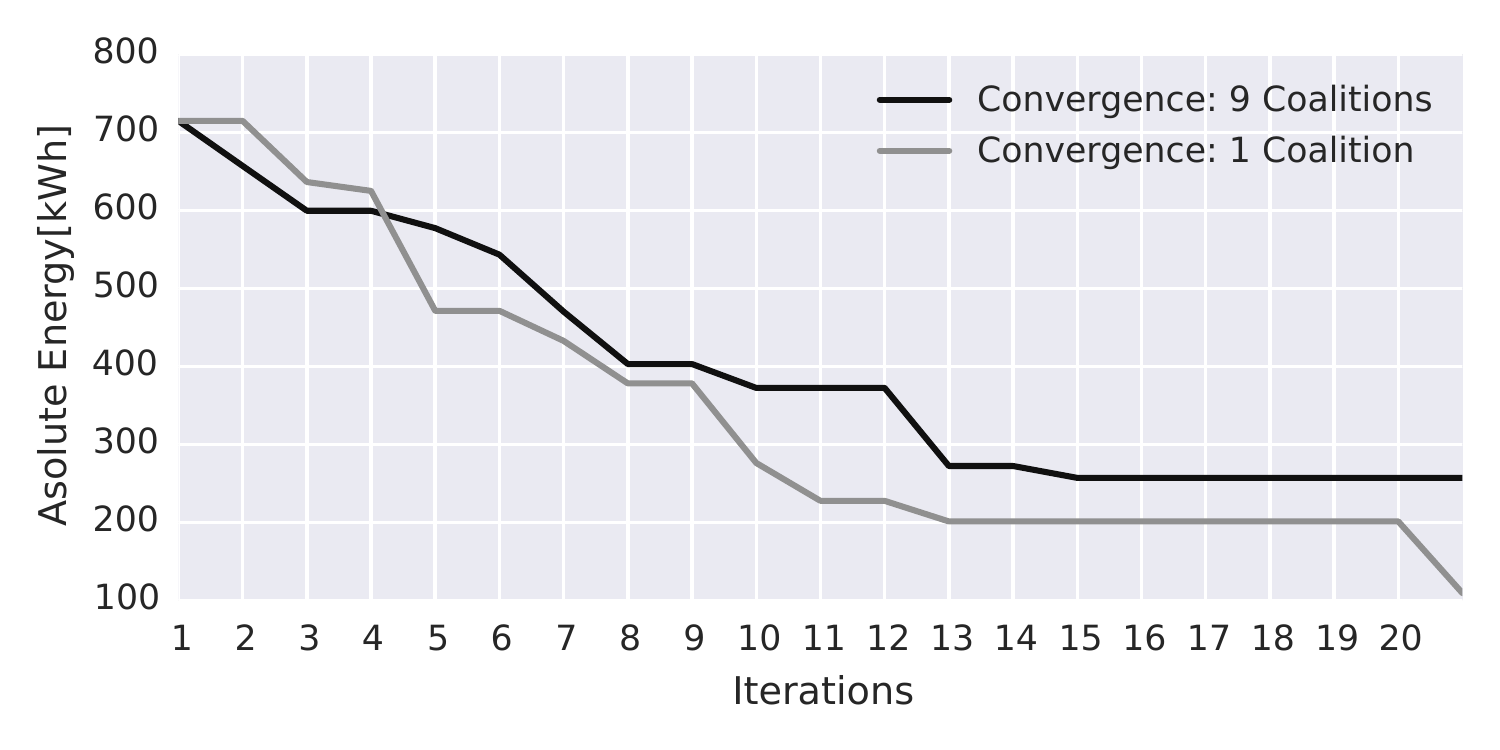}
\caption{Convergence as minimization of energy interaction with \emph{Utility}}
\label{graph_convergence}
\end{figure}



\begin{figure}[t]
\centering
\includegraphics[scale=0.35]{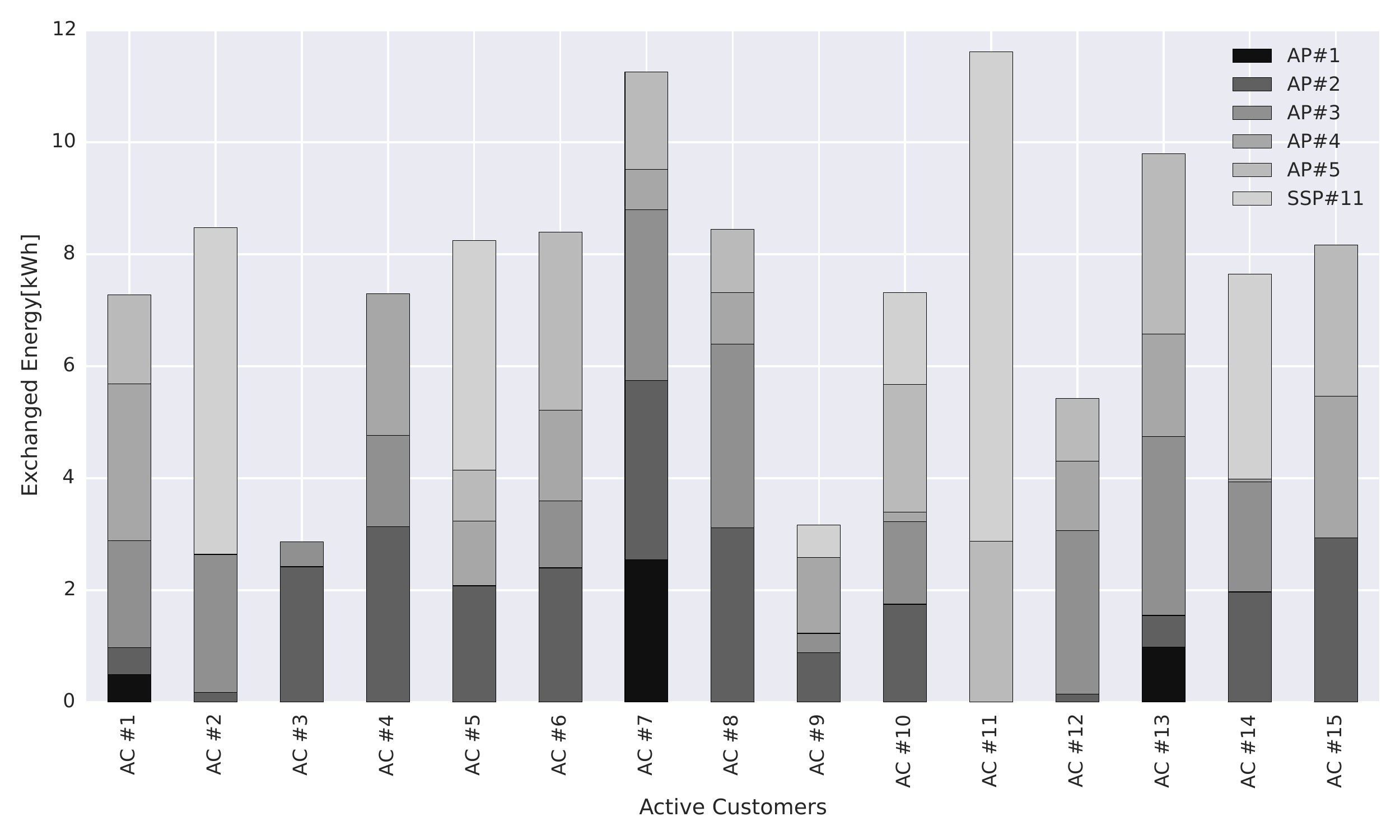}
\caption{Local and distributed energy exchange within APs and ACs (for SSP\#2)}
\label{graph_matching}
\end{figure}

\begin{figure}[h]
\centering
\includegraphics[width=85mm, height=45mm]{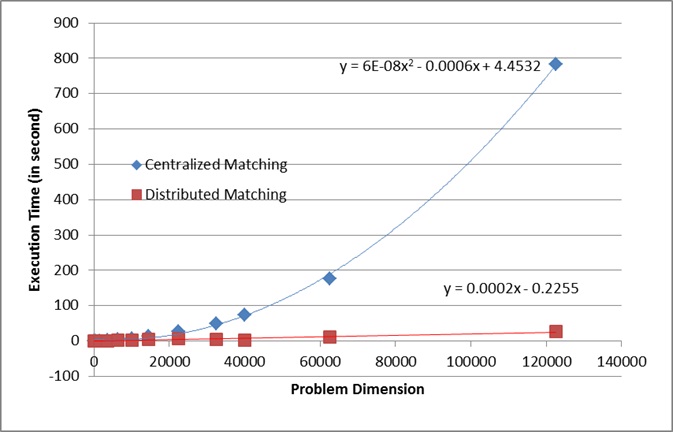}
\caption{Comparison of execution time within Centralized and Distributed Matching schemes}
\label{graph_et_comparison}
\end{figure}

\subsection{1st Study}
The 1st study considers the effect of coalitions over a hypothetical distribution system containing no Passive customers. For this purpose, a distribution system with 20 SSPs (operating under 1 ESP), each having 10 consumers and 5 producers, is considered. The energy statuses of these 15 customers (10+5) are randomly generated. For example, at a particular matching instance, the total demand in \emph{SSP\#2} is \emph{127.1} kWh (distributed over 10 consumers) and total supply in \emph{SSP\#2} is \emph{76.1} kWh (distributed over 5 producers). Therefore, \emph{for SSP\#2}, the energy status is \emph{-51} kWh (saying, it will require 51 kWh from \emph{Utility}). 

The coalition formation engine produces 9 coalitions out of 20 SSPs. For example, SSPs \emph{16, 2, 11, 6, and 7} form a coalition (provided by the ANM, considering energy status profile of all SSPs at a certain time period). So, they will exchange energy within themselves in addition to their own local exchange. The comparison of (absolute) energy status (essentially the \emph{Utility} interactions, before and after performing distributed matching operation) is shown in Figure~\ref{graph_e_status}. For comparison, we also showed the meshed interactions (i.e. 1 coalition). The 9 coalition case cannot reach the ultimate convergence since it used only limited communication network. The matching operation, as expected, reduces the energy interaction with \emph{Utility}. Coming back to \emph{SSP\#2}, the energy interaction is reduced with the \emph{Utility}. Interestingly, for some of the SSPs (e.g. \emph{SSP\#11}, the matching operation using 9 coalition completely reduces \emph{Utility} interactions while using 1 coalition still decides to sell back \emph{22.5} kWh to \emph{Utility}. 

The \emph{iteration} in this context is basically recorded when one of the SSPs settles on its local and distributed exchange in the respective scopes. The reduction pattern of absolute energy status accumulated over all SSPs (i.e. the \emph{Utility} interactions for the ESP) with each \emph{iteration} is presented in Figure~\ref{graph_convergence} for both of the cases. In brief, the \emph{Utility} interaction is reduced from \emph{714.6} kWh to \emph{257} kWh (for 9 coalitions) and to \emph{109} kWh for 1 coalition. We have chosen \emph{SSP\#2} to detail the distributed energy matching operation. For the time being, all the consumers and producers are considered as AC and AP, respectively (by setting the variable $bounds(j)$ as $0$). The incorporation of Passive customer is thus straight forward. The energy exchange within the APs and ACs (inside \emph{SSP\#2}) as well as interaction of ACs with \emph{SSP\#11} is pointed in Figure~\ref{graph_matching}. For example, \emph{AC\#2} of \emph{SSP\#2} will receive \emph{5.83} kWh from \emph{SSP\#11}. Note that, only \emph{SSP\#11} participates in the energy transaction. Therefore, the proposed distributed algorithm (Algorithm~\ref{dist_e_matching}) ensures that a minimum number of external SSPs will be involved in the transaction. The distributed matching takes 21 \emph{iterations} to converge. 

\begin{figure}[h]
\centering
	\includegraphics[scale=0.35]{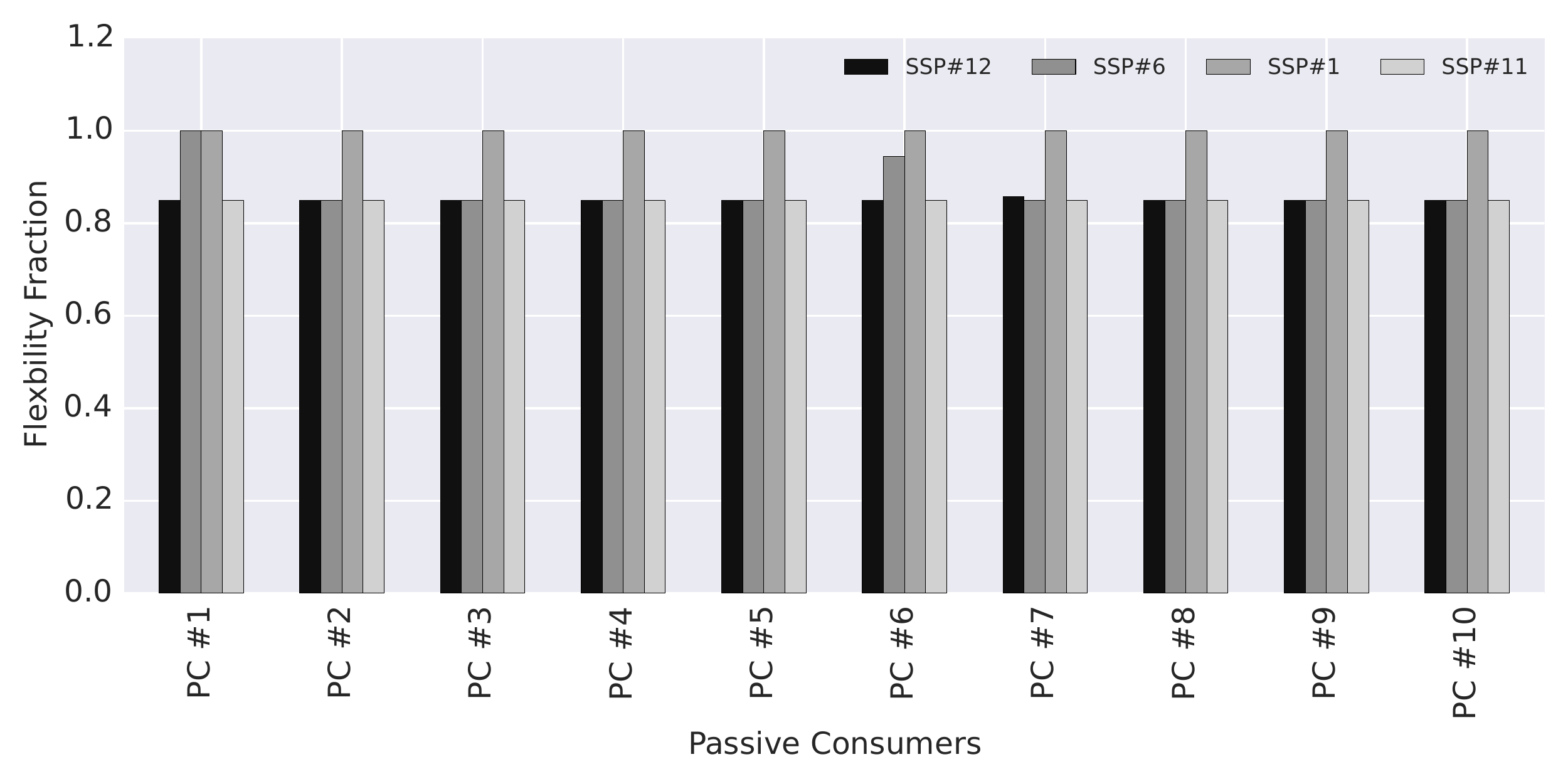}
	\caption{Flexibility of PCs}
	\label{flex_pc}
\end{figure}

\begin{figure}[h]
\centering
	\includegraphics[scale=0.35]{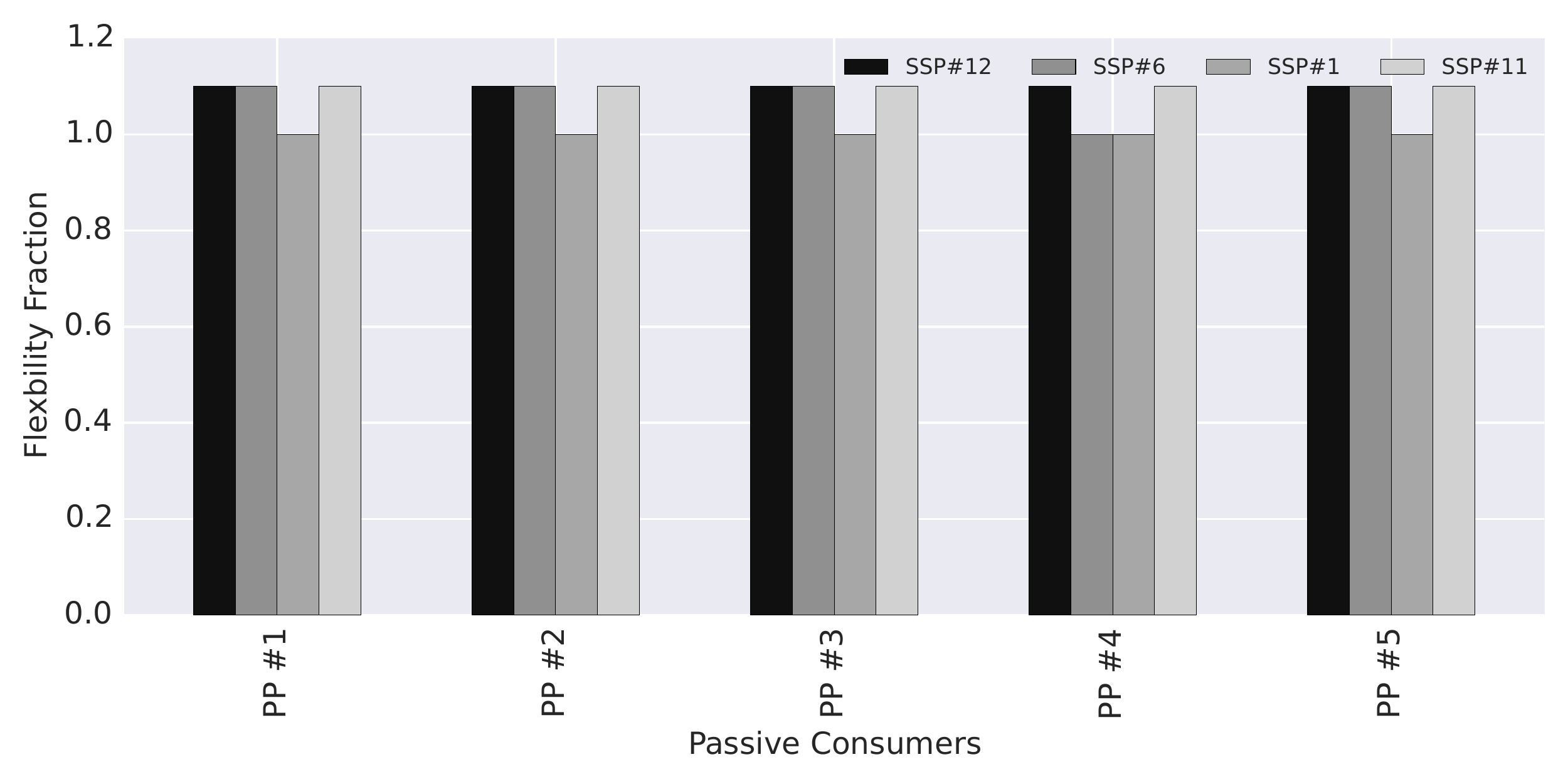}
	\caption{Flexibility of PPs}
	\label{flex_pp}
\end{figure}


One of the advantages of incorporating distributed energy matching is the reduced algorithmic complexity. The complexity (consequently, the execution time) of centralized matching increases exponentially with the number of customers. Therefore, we have presented a comparative study showing the execution times in centralized and distributed matching (1 coalition) over the dimension of the problem. The dimension is defined as \scriptsize$|SSPs|\times|ACs/SSP|\times|APs/SSPs|$\normalsize. The execution time comparison with centralized matching (i.e. $|SSPs|$ is 1) with distributed matching (considering Case 3, $|SSPs|$ is 10) is shown in Figure~\ref{graph_et_comparison}. The distributed matching scheme dominates significantly (computationally) over its centralized counterpart for large and diverse distribution system.


\begin{figure}[h]
\centering
\includegraphics[scale=0.25]{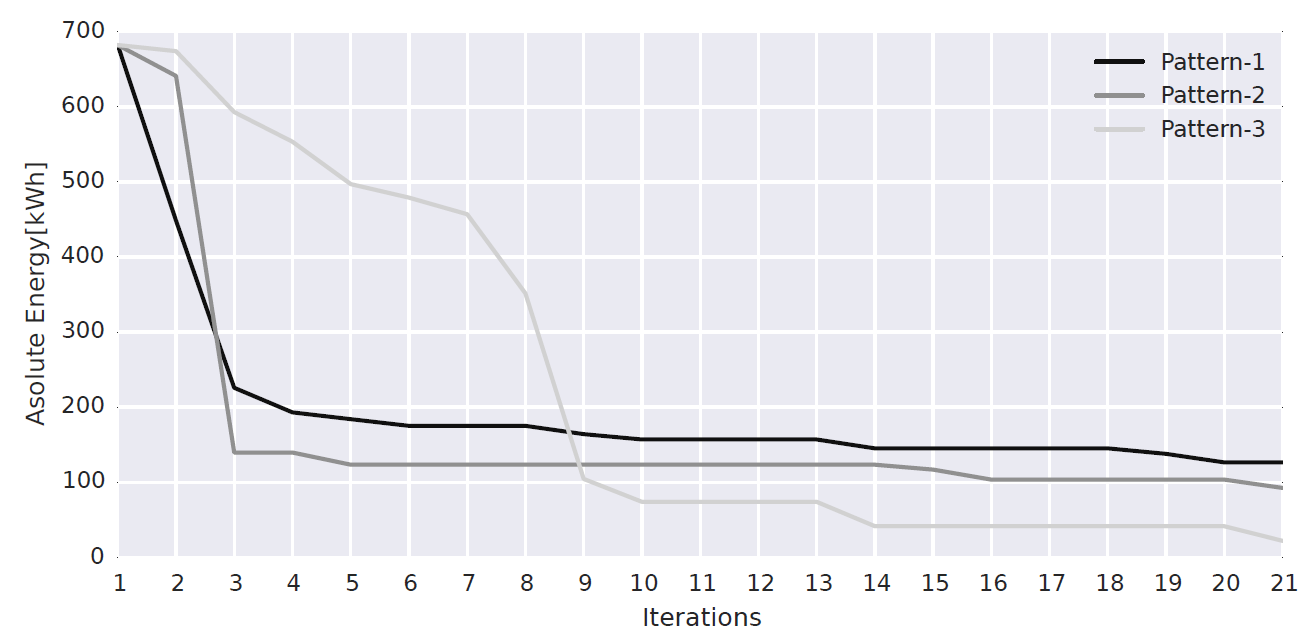}
\caption{Effect of increasing number of passive subscribers}
\label{passive_effect}
\end{figure}

\subsection{2nd Study}
The 2nd study considers an increasing number of subscribers (35 consumers and 10 producers) for each of the 20 SSPs. Among 35 consumers, 10 are assumed to be Passive Consumer (PC) with maximum flexibility of 15\% and the remaining 25 are Active Consumers (AC). At the same time, among 10 producers, 5 are assumed to be Passive Producers (PP) with maximum flexibility of 10\% while the remaining 5 are assumed as Active Producers (AP). For the sake of simplicity, in this case, the coalition scheme is avoided (unlike the 1st study).

The flexibility of Passive Consumers (PC) for four SSPs are provided in Figure~\ref{flex_pc}. As mentioned before, the flexibility of a PC determines the reduction fraction of the energy consumption. The figure points out the flexibility status of PCs. For example, PC\#6 of SSP\#6 needs to reduce its energy down by 6\% in order to maximize the local energy matching. In other words, PC\#6, which can reduce its energy maximum of 15\% due to the event such as DR, will only need to reduce 6\% because of the distributed optimization process. On the other hand, SSP\#1 does not need any of its PCs to reduce down the consumption (unlike SSP\#11, where all of its PC need to reduce down to the maximum fraction). Similarly, Figure~\ref{flex_pp} shows the flexibility of different Passive Producers (PPs).

The effect of increasing number of passive subscribers on the distributed matching process can be seen in Figure~\ref{passive_effect}. The Pattern-1 describes the convergence pattern (measured by the absolute energy transactions with \emph{Utility}) of the 2nd study while Pattern-2 depicts the effect when the number of passive subscribers is increased. For Pattern-2, number of PP is increased from 5 to 7 and number of PC is increased from 10 to 20 while keeping the total number of subscribers same. As seen in Figure~\ref{passive_effect}, the increase in passive subscribers essentially increases the accumulate flexibility of the matching process and hence producing a better convergence graph than that of in 2nd study. Specifically speaking, increasing the number of passive subscriber as mentioned before reduces the \emph{Utility} interactions from roughly \emph{126}kWh to \emph{92}kWh by utilizing the consequent increased flexibility. The absolute energy transaction goes further down (to \emph{26}kWh) when all of the subscribers are passive (that consequently further increases the flexibility of the system). The phenomena is plotted as Pattern-3 in Figure~\ref{passive_effect}.

\section{Conclusion}
\label{conclusion}

We introduce a secured and distributed demand side management framework for future smart distribution grid taking the advantage of flexible power exchange architecture (such as Digital-Grid \cite{Abe:2011}, and ``ECO Net''\cite{2003}). The framework is particularly designed for new market entrant (e.g. Power Producer and Supplier, \emph{PPS}) operating on deregulated energy market in Japan. Along the way, we have proposed a distributed matching algorithm that decides on a scheduled commitment (determined $N$-time ahead) of energy exchange to be followed in real-time. We have identified the boundaries and roles of potential players who can be benefited from the proposed matching scheme. The contributions of the article are two-fold: 1) the energy matching service design for the future electricity market by strategically grouping subscribers based on their ability to commit to a certain energy profile, 2) the secured distributed matching operation considering a very high number of subscribers, their preferences and commitments. Therefore, the proposed framework aligns with the PPS's business model. A \emph{learning based coalition formation} method for adaptive network mapping is also proposed that essentially provides the reduced communicative network for energy exchange within the players. As of now, we have showed the day-ahead distributed energy matching scheme. The security and privacy of each subscriber are maintained by avoiding exposition of peer energy information (down to subscriber level) and by ensuring minimized and aggregated energy interactions within aggregators (i.e. SSPs). Therefore, the proposed framework aligns perfectly with the security and privacy requirement from relatively conservative electricity market. Although, the proposed framework focuses on Japanese power market, the framework can be effectively utilized in other markets as well by making appropriate assumptions. The article is particularly limited to the off-line ($N$-time ahead) energy commitment amongst subscribers, $SSP$ (microgrids) and \emph{Utility}. The design of pricing scheme is not provided in the manuscript. However, the pricing schemes assumed to be \emph{incentive compatible}. 


\end{document}